# Engineering the work function of armchair graphene nanoribbons using strain and functional species: a first principles study

Xihong Peng,[1,*] Fu Tang[2], and Andrew Copple[2]

[1] Department of Applied Sciences and Mathematics, Arizona State University, Mesa, AZ 85212

[2] Department of Physics, Arizona State University, Tempe, AZ 85287

## ABSTRACT

First principles density-functional theory calculations were performed to study the effects of strain, edge passivation, and surface functional species on the structural and electronic properties of armchair graphene nanoribbons (AGNRs) with a particular focus on the work function. The work function was found to increase with uniaxial tensile strain while decreasing with compression. The variation of the work function under strain is primarily due to the shift of the Fermi energy with strain. In addition, the relationship between the work function variation and the core level shift with strain is discussed. Distinct trends of the core level shift under tensile and compressive strain were discovered. For AGNRs with the edge carbon atoms passivated by oxygen, the work function is higher than for nanoribbons with the edge passivated by hydrogen under a moderate strain. The difference between the work functions in these two edge passivations is enlarged (reduced) under a sufficient tensile (compressive) strain. This has been correlated to a direct-indirect band gap transition for tensile strains of about 4% and to a structural transformation for large compressive strains at about -12%. Furthermore, the effect of the surface species decoration, such as H, F, or OH with different covering density, was investigated. It was found that the work function varies with the type and coverage of surface functional species. F and OH decoration increase the work function while H decreases it. The surface functional species were decorated on either one side or both sides of AGNRs. The difference in the work functions between one-side and two-side decorations was found to be relatively small, which may suggest an introduced surface dipole plays a minor role.

I.  Introduction

Graphene is a single atomic layer carbon sheet in a honeycomb lattice. Due to its exceptionally high crystalline quality, graphene demonstrates a unique linear dispersion relation and the charge carriers behave as massless fermions [1]. Experiments on graphene have shown the charge mobility exceeded 15,000 $cm^2/(Vs)$ even under ambient conditions [2]. Graphene has been considered as a promising material for many advanced applications in future electronics [1, 2]. Engineering of the structure and electronic properties of graphene is essential for these applications. Recently, tunability of the work function has drawn a particular attention [3-14]. For example, in an electronic device using graphene as an active channel layer, the work function of graphene determines the band alignment [4] and directly affects the charge injection between graphene and metallic contact [5, 6]. Graphene is also considered as transparent electrode [4, 7] and cathode materials [8] in optoelectronic devices. The work function will be critical for maximized energy conversion efficiency. Its atomically thin nature makes vertically standing graphene a promising candidate as a field emitter. [9] A lower work function can dramatically enhance the emitting current [9]. Different approaches have been investigated to modulate the work function, such as employing an external electric field [10], chemical [4] and metal doping [5, 11], substrate orientation [3, 12], and a self-assembled monolayer [13].

As a practical issue, strain is almost inevitable in fabricated graphene structures, manifesting as the formation of ridges and buckling [15, 16]. Graphene possesses superior mechanical stability. It can sustain a tensile strain up to 30% demonstrated by Kim *et al* [17]. A number of studies have investigated the effect of strain on the electronic properties of graphene and graphene nanoribbons [18-25], such as the band gap and mobility. To advance the graphene based technology, it will be essential to examine how strain affects the work function of graphene and graphene nanoribbons. In addition, functional group decoration/passivation is another factor that can be practically involved in the preparation process of graphene. In fact, an extensive effort has been made to employ functional species for tailoring the properties of graphene [24, 26-32]. The combined effect of strain and decoration/passivation on the work function of graphene is also of great interest to investigate.

In the present work, first principles density-functional theory [33] calculations were conducted to investigate the work function of edge passivated armchair graphene nanoribbons (AGNRs) modulated by external uniaxial strains and surface species decoration. Two groups of edge passivation (H and O) and three types of surfaces species (H, F, and OH) have been studied. It has been demonstrated that the strain can effectively tune the work function of graphene nanoribbons by primarily shifting the Fermi level. Sufficient strain on edge-O passivated AGNRs yields a structural or direct-indirect band gap transition, which can produce a significant change in the work function. Furthermore, it was found that the work function varies with the type and coverage of surface function group. Surface dipoles here have less effect on the work function compared with the surface states introduced by the functional group.

## II. Computational details

First principles density-functional theory [33] calculations were carried out using Vienna *ab-initio* Simulation Package [34, 35]. The local density approximation (LDA) [33] was applied. The generalized gradient approximation (GGA) [36] was also used to check the work function of graphene. It was found that the work function predicted by LDA and GGA is 4.48 eV and 4.49 eV, respectively. Both values are in good agreement with other theoretical [5, 18] and experimental [37] studies. The pseudo-potential plane wave approach was employed. The kinetic energy cutoff of the plane wave basis was set to be 450.0 eV. Core electrons of atoms were described using Vanderbilt ultra-soft pseudo-potentials [38]. The reciprocal space was sampled using $4 \times 1 \times 1$ Monkhorst-Pack grid [39] centered at the Γ point. A total of 21 K-points were included in the band structure calculations along Γ (0, 0, 0) to X (0.5, 0, 0). The Gaussian smearing method was used to describe partial occupancies of orbitals, with width of the smearing set at 0.05 eV. The total energy in the self-consistent scheme was converged to within 0.01 meV. Atoms were fully relaxed until force and stress components are less than 0.02 eV/Å and 1.0 kbar, respectively. The initial lattice constant along the armchair direction (i.e. *x*-axis) in a ribbon was set to 4.22 Å, taken from the 2D graphene sheet. The lattice constant of all AGNRs was fully optimized through the technique of energy minimization. The vacuum distance between the ribbon and its replicas is about 30 Å (*y* direction) and 16 Å (*z* direction) to eliminate interaction between ribbon replicas due to periodic boundary condition.

## III. Results and discussion

### A. Strain modulated AGNRs

We first studied the effect of uniaxial strain on the electronic properties such as the band structure, work function, and core level shift of the AGNRs. Two types of edge passivation, H and O, were studied, as shown in Figs. 1(a) and 1(b). The width of an AGNR is measured as the distance between two carbon atoms on both edges, which is related to the number of C atoms along the zigzag direction (i.e. the *y* direction in Fig.1). It is known that, due to quantum confinement effects, AGNRs can be classified into three families according to the width of the AGNR in which the number of C atoms in the zigzag direction falling in the categories of $3n$, $3n+1$, and $3n+2$, where n is a positive integer [40, 41]. In this work, three widths of AGNRs were chosen, 13.4 Å, 14.6 Å, and 15.8 Å, corresponding to the number of C atoms of 12, 13, and 14, respectively, in the zigzag direction. Shown in Figs. 1(a) and 1(b) are AGNRs with a width of 14 C atoms. We also investigated AGNRs with widths of 12 and 13 C atoms, representing other two families.

The structures of the ribbons were fully optimized through energy minimization. Based on the relaxed structure with an optimized lattice constant, uniaxial strain within the range of ±16% was applied by scaling the lattice constant. The positive values of strain refer to uniaxial tensile strain, while negative values correspond to compression. Note that the *y* and *z* coordinates of the ribbon are further relaxed at a given strain.

### A.1) Work function and core level shift

The work function of an AGNR is defined as the energy difference between the vacuum and Fermi levels,

$$\phi = V_{vacuum} - E_{Fermi} \tag{1}$$

In numerical calculations, the Fermi level $E_{Fermi}$ is determined by integrating the density of states from the lowest energy level to an energy level (i.e. Fermi energy) which gives a total number of electrons in the unit cell. Specifically, in this work, the Fermi level of the semiconducting AGNRs was set to be the middle of the band gap. The vacuum potential $V_{vacuum}$ is read from the plot of planar-average electrostatic potential energy along the *z* direction (i.e. pick the value in the middle of vacuum from the plot). All electronic energies of a ribbon in this study are referenced to its vacuum potential energy.

The work function of the AGNRs was studied as a function of uniaxial strain. Both edge-H and edge-O passivation were investigated for AGNRs with different widths. As an example, the strain dependence of the work function in the AGNRs with a width of 14 C atoms is plotted in Fig. 2. It was found that the work function increases with tensile strain and decreases with compressive strain. This observation is similar to the result obtained with a strained 2D graphene sheet [18].

Since the work function is determined by two energy levels, $E_{Fermi}$ and $V_{vacuum}$, it is worth examining the strain dependence of these two terms. Fig. 3(a) shows the Fermi and vacuum levels as a function of strain for the edge-H passivated AGNR. It shows that the strain has a dominating effect in shifting the Fermi level while only having a minimal effect on the vacuum potential energy. For example, the change of the Fermi energy within the strain range considered in this study is 1.59 eV, while the variation of the vacuum level is only 0.13 eV, which suggests that the variation of the work function is mainly contributed by the shift of the Fermi level.

As shown below, the variation of the work function is also correlated to the core level shift of carbon atoms in the AGNR. To see this, the work function is rewritten as the following formula [42, 43] by inserting the average electrostatic potential energy at carbon ionic cores $V_{core}$,

$$\phi = (V_{core} - E_{Fermi}) - (V_{core} - V_{vacuum}) \tag{2}$$

The first term is used for determining the core level shift in a solid film. The second term corresponds to the core level shift relative to the vacuum level.

Both terms in Equation 2 can be referenced to their values of the relaxed AGNR and plotted in Fig. 3(b), where

$$\Delta(V_{core} - E_{Fermi}) = (V_{core} - E_{Fermi})_\varepsilon - (V_{core} - E_{Fermi})_0 \tag{3}$$

$$\Delta(V_{core} - V_{vacuum}) = (V_{core} - V_{vacuum})_\varepsilon - (V_{core} - V_{vacuum})_0 \tag{4}$$

The referenced core level shift relative to the vacuum level in Equation 4 decreases (increases) with tensile (compressive) strain. Its variation with strain is nearly linear. The change of this shift with strain is mainly due to the electrostatic potential variation with the modulated distance between ionic cores and valence electrons [44]. When the ribbon is under a tensile strain, the valence electrons are further apart from the ionic cores, so the electrostatic potential contributed by valence electrons to the ionic cores is reduced. This causes the potential energy of the ionic cores to decrease.

The core level shift relative to the Fermi energy in Equation 3, however, demonstrates a different behavior with strain, shown in Fig. 3(b). From the curve, we can see that this shift is reduced significantly to a smaller value in the compressed strain while the tensile strain barely affects the value. Here, the change in the electrostatic potential due to the variation of the distance between the ionic cores and valence electrons may have been dominated (nearly canceled) by the shift of the Fermi energy in the compressed (expanded) AGNR.

In addition, we studied the AGNRs with different widths, such as 12 and 13 carbon atoms in the zigzag direction. The general trends presented in Fig. 2 and Fig. 3 are also valid for those widths. The distinct trend of core level shift relative to the Fermi energy under tensile and compressive strain make the traditional electron spectroscopy tools such as X-ray photoelectron spectroscopy valuable for characterizing the strain in graphene [45, 46]. This is attractive practically since the strain can be easily introduced into the monolayer graphene structures during preparation processes.

In Figure 2, it is also found that the work function of the edge-O passivated AGNRs is higher than that of the edge-H passivated nanoribbons under a moderate strain. At a strain larger than ~ 4%, the difference in the work function between these two types of edge passivation increases, while this difference starts to reduce under a compressive strain (~ -12%). As shown below, the deviation of the work function trend under large compressive/tensile strains is correlated to structural/electronic transition of the edge-O passivated AGNRs.

### A.2) Structure and band gap transition in edge-O passivated AGNRs

It is interesting to observe electronic and structural transitions in the edge-O passivated AGNRs under large uniaxial tensile and compressive strain, respectively. When a large tensile strain is applied, the band gap of the AGNRs shrinks to zero [24]. For example, the ribbon with a width of 14 C atoms demonstrates a zero-gap at +8% strain. In order to closely examine the gap variation with strain, the band structures of the edge-O passivated AGNR under different values of strain are presented in Fig. 4. Under +4% strain, the band gap experiences a transition from direct to indirect. With increasing tensile strain, the indirect gap decreases to zero at +8% strain. Beyond +8% strain, no gap exists. This electronic evolution with strain is mainly due to edge defects produced by tensile strain (for a detailed discussion please refer to reference [24]). The direct-indirect gap transition is also related to the significantly higher work function of the edge-O AGNR at large tensile strain (see Fig. 2).

On the other hand, the band structure of the AGNR under -12% strain is largely deviated from the relaxed one. It was found that a structural transformation occurs at this strain and larger compression. To illustrate the structure transition, the geometries of the relaxed and -12% strained AGNRs are shown in Figs. 5(a) and 5(b). The pentagon formed by the edge O and four neighboring C in Fig. 5(a) transfers to a heptagon under -12% strain in Fig. 5(b). And the hexagon formed by six C atoms (labeled 2, 4, 6, 1', 3', and 5' in Fig. 5(a)) transforms to a quadrilateral in Fig. 5(b). Besides these rearrangements near the edges of the ribbon, the bond lengths of horizontal carbon pairs (i.e. C3-C4, C5-C6, C9-C10, C11-C12, C13-C14, C15-C16, etc.) are distinct in the relaxed and -12% strained AGNRs. For example, in the relaxed ribbon, the pairs of C3-C4, C9-C10, and C13-C14 form the C-C bond with bond lengths ~1.4 Å, while pairs of C5-C6, C11-C12, and C15-C16 do not form bonds with C-C distances ~ 2.7 Å. However, it is opposite in the -12% strained ribbon, where the latter pairs all form C-C bonds while the former pairs are apart, shown in Fig. 5(b). The detailed bond lengths of these two structures are listed in Table 1. A similar structure transformation was also found at the -12% strain in the edge-O passivated ribbon with widths of 12 and 13 C atoms.

### B. Surface functional species decorated AGNRs

Recently tailoring the properties of graphene by surface functional groups has attracted a tremendous interest [24, 26-32]. Here we investigate effects of three types of surfaces species (H, F, and OH) on structural and electron properties of AGNRs.

### B.1) Structural properties

The starting ribbon is a geometrically relaxed edge-H passivated AGNR with a width of 14 C atoms, as shown in Fig. 6(a). Based on this ribbon, different surface species, such as H, F, and OH, is decorated on the ribbon surface on either one side or both sides. The number of the decorating surface species varies as 2, 4, 6 and 8, with each addition of one surface atom corresponding to an increment of 3.57% surface coverage (i.e. 1/28, 28 is the number of carbon atoms in the unit cell). For example, Fig. 6(b) and 6(c) show the geometrically relaxed ribbons with four and eight H atoms (represented by 4H and 8H) decorated on one side of the ribbon, respectively. It is clear that the ribbons were bent for these cases due to the lattice distortion by the surface decorated species on the same side [26]. The bent geometry compromises the local stress induced by the lattice distortion. Fig. 6(d) shows the relaxed structure of eight H atoms on both sides of the ribbon (4 H on each side). The ribbon is not bent and nearly planar. For other species, F and OH, the relaxed ribbons are similar to those shown in Fig. 6(b) - 6(d).

For all AGNRs with different surface species, their structures were fully relaxed through energy minimization. The lattice constant of the AGNRs were also optimized so that the force and stress components on each atom were converged to within 0.02 eV/Å and 1.0 kbar, respectively. The optimized lattice constants were reported in Fig. 7(a). It shows that the edge-O passivated ribbon without surface species has the shortest lattice constant of 4.10 Å. For the edge-H passivated ribbon, the lattice constant increases to 4.25 Å. The lattice constant slightly increases with the number of surface H species decorated on the ribbon, to 4.30 Å with eight surface H atoms. A similar effect was also found in the surface species F and OH. In addition, for the same number of surface species, there is no significant difference in the lattice constant between the cases of one-side and two-side decoration.

As shown in Fig. 6(b) and 6(c), the relaxed ribbons are bent if the species were decorated on one side of the ribbon. Increasing the density of the surface species will increase the ribbon bending curvature (defined as $1/R$, where $R$ is the radius of the bending ribbon). The bending curvature of the AGNRs as a function of surface species coverage is plotted in Fig. 7(b). It is clear that the curvature increases rapidly with increasing number of surface species. For example, the curvature of the ribbon with 4H and 8H surface species are 0.038 Å$^{-1}$ and 0.203 Å$^{-1}$, respectively, which have corresponding curvature radii of 26.12 Å and 4.92 Å, respectively. In addition, the AGNRs with widths of 12 and 13 C atoms in the zigzag direction were explored. Our calculations showed that there is no distinct difference in the predicted curvature for the AGNRs with different widths. Our calculated bending structures and curvatures of AGNRs were consistent with that of first-principles molecular dynamics simulation conducted by Yu and Liu [26].

### B.2) Band structures and density of state

Electronic properties were investigated for the AGNRs with different surface species. As an example, band structures and the corresponding density of states (DOS) for the AGNRs with one-side surface species decoration are plotted in Fig. 8. As a reference, the band structure and DOS for the starting ribbon (i.e. the edge-H passivated ribbon without surface species) is also presented in Fig. 8(a). All energies are referenced to the vacuum level. The Fermi level is represented by the horizontal dashed line. It was known that the edge-H passivated AGNR is a semiconductor, shown in Fig. 8(a). However, introduction of the surface species, such as H, F and OH, bring surface states near the Fermi level. Increasing the coverage density of the surface species, the band structures and DOS are deviated farther from that of the starting ribbon. For example, Figs. 8(b), 8(f) and 8(j) display the band structures and DOS for 2H, 2F, and 2OH surface species, respectively. Two surface states were brought in near the Fermi level. However, the energy bands and DOS in which energies are far away from the Fermi level are similar to that of Fig. 8(a). Increasing the number of surface species from two to four, more surface states were shown near the Fermi level, and the DOS is further modified, as shown in Figs. 8(c), 8(g), and 8(k). Continuing to increase the number of the surface species to six and eight, the band structure and DOS were significantly modified.

We also studied the band structures and DOS for the AGNRs with surface species decorated on both sides of the ribbon. It is interesting to note that the band structures and DOS are very similar to those of one-side decoration. For example, the band structure and DOS for the AGNR with each side decorated by one H atom is very close to the one in Fig. 8(b).

### B.3) Work function

The calculated work function of the AGNRs is reported as a function of the number of surface species in Fig. 9. Generally, different surface species affect the work function in a different manner. For example, adding surface H to the ribbon decreases its work function, while adding surface F or OH increases the work function. Therefore, the surface species can be classified into two groups, one increasing the work function (such as F and OH), and the other reducing it (such as H).

The work function shift by the surface decoration could come from two sources: (i) molecular dipoles formed between the decorated species and the ribbon surface; (ii) charge rearrangements induced by the chemical bond formation between the decorated species and the ribbon surface [47-49]. We argued that the work function shift here is primarily due to the latter. This can be seen from the surface states introduced near the Fermi level in the band structures (Fig. 8). These surface states come from the transition of carbon atoms from $sp^2$ to $sp^3$ hybridization [30]. F and OH with a higher electronegativity could introduce deeper states while H with a lower electronegativity could introduce shallower states. For example, the surface bands introduced by 2F and 2OH surface species are in the range of -4.5 ~ -5.5 eV, and -4.2 ~ -5.0 eV, respectively (see Figs. 8(f) and 8(j)). However, the surface band introduced by 2H species is in the range of -3.9 ~ -4.2 eV, which is much shallower, shown in Fig. 8(b).

The contribution of the molecular dipole from surface decoration can be evaluated from the work function difference between one side and two side decoration. In one side decoration, the molecular dipole will introduce an additional potential change to the vacuum level [47, 50], while in the two-side decoration, molecular dipoles on the two faces will tend to cancel each other. From Figure 9, it is found that work function difference between one side and two side decorations is small, which suggests the introduced surface dipole plays a minor role here. The molecular dipole $C^+$-$F^-$/$C^+$-OH enhances the potential barrier to the vacuum level [47, 50] For the H group, however, the molecular dipole $C^-$-$H^+$ reduces this potential barrier. Another possible factor contributing to the work function difference between one side and two side decoration is the deviation in the ribbon geometries, which indicates the existence of the different local strain at these two cases. For example, for the 8H surface species, the two-side decoration yields a nearly planar ribbon, while the one-side decoration produces a largely bent structure, shown in Figs. 6(d) and 6(c).

### IV.  Summary and conclusion

In summary, using first principles density-functional theory calculations, it was found that (1) the work function of AGNRs increases with tensile strain, and decreases with compressive strain, regardless of the type of edge passivation O and H; (2) the core level shift relative to the Fermi energy decreases with compressive strain, while tensile strain only affect it slightly; (3) the edge-O passivated AGNRs experiences a direct-to-indirect band gap transition under sufficient tensile strain and a structural transformation occurs with a large compressive strain; (4)  F and OH surface decoration increases the work function while H decoration decreases the work function of AGNRs; (5) one-side and two-side surface species decoration brings only relatively small difference in the work function, given the same number of surface species.

**Acknowledgement**

This work is supported by the Research Initiative Fund from Arizona State University (ASU) to Peng. The authors thank ASU Advanced Computing Center (Saguaro) for providing computational resources and Nathaniel Ralston for the critical review and proofreading of the manuscript.

* Author to whom correspondence should be addressed.  Electronic mail: xihong.peng@asu.edu.

**Reference:**

[1]     Novoselov K S, Geim A K, Morozov S V, Jiang D, Katsnelson M I, Grigorieva I V, Dubonos S V and Firsov A A 2005 *Nature* **438** 197
[2]     Novoselov K S, Geim A K, Morozov S V, Jiang D, Zhang Y, Dubonos S V, Grigorieva I V and Firsov A A 2004 *Science* **306** 666
[3]     Wang B, Gunther S, Wintterlin J and Bocquet M L 2010 *New Journal of Physics* **12** 043041
[4]     Shi Y, Kim K K, Reina A, Hofmann M, Li L-J and Kong J 2010 *ACS Nano* **4** 2689
[5]     Giovannetti G, Khomyakov P A, Brocks G, Karpan V M, van den Brink J and Kelly P J 2008 *Phys. Rev. Lett.* **101** 026803
[6]     Khomyakov P A, Giovannetti G, Rusu P C, Brocks G, van den Brink J and Kelly P J 2009 *Phys. Rev. B* **79** 195425
[7]     Jo G, Na S I, Oh S H, Lee S, Kim T S, Wang G, Choe M, Park W, Yoon J, Kim D Y, Kahng Y H and Lee T 2010 *Appl. Phys. Lett.* **97** 213301
[8]     Cox M, Gorodetsky A, Kim B, Kim K S, Jia Z, Kim P, Nuckolls C and Kymissis I 2011 *Appl. Phys. Lett.* **98** 123303
[9]     Eda G, Unalan H E, Rupesinghe N, Amaratunga G A J and Chhowalla M 2008 *Appl. Phys. Lett.* **93** 233502
[10]    Yu Y-J, Zhao Y, Ryu S, Brus L E, Kim K S and Kim P 2009 *Nano Lett.* **9** 3430
[11]    Benayad A, Shin H-J, Park H K, Yoon S-M, Kim K K, Jin M H, Jeong H-K, Lee J C, Choi J-Y and Lee Y H 2009 *Chemical Physics Letters* **475** 91
[12]    Murata Y, Starodub E, Kappes B B, Ciobanu C V, Bartelt N C, McCarty K F and Kodambaka S 2010 *Appl. Phys. Lett.* **97** 143114
[13]    Park J, Lee W H, Huh S, Sim S H, Kim S B, Cho K, Hong B H and Kim K S 2011 *The Journal of Physical Chemistry Letters* **2** 841
[14]    Yi Y, Choi W M, Kim Y H, Kim J W and Kang S J 2011 *Appl. Phys. Lett.* **98** 013505
[15]    Cai Y Q, Zhang A H, Feng Y P, Zhang C, Teoh H F and Ho G W 2009 *J. Chem. Phys.* **131** 224701
[16]    Wang Y D, Lu N, Xu H B, Shi G, Xu M J, Lin X W, Li H B, Wang W T, Qi D P, Lu Y Q and Chi L F 2010 *Nano Res.* **3** 520
[17]    Kim K S, Zhao Y, Jang H, Lee S Y, Kim J M, Kim K S, Ahn J-H, Kim P, Choi J-Y and Hong B H 2009 *Nature* **457** 706
[18]    Choi S-M, Jhi S-H and Son Y-W 2010 *Phys. Rev. B* **81** 081407
[19]    Gui G, Li J and Zhong J X 2008 *Phys. Rev. B* **78** 075435
[20]    Li Y, Jiang X W, Liu Z F and Liu Z R 2010 *Nano Res.* **3** 545
[21]    Lu Y and Guo J 2010 *Nano Res.* **3** 189
[22]    Pereira V M and Neto A H C 2009 *Phys. Rev. Lett.* **103** 046801
[23]    Sun L, Li Q X, Ren H, Su H B, Shi Q W and Yang J L 2008 *J. Chem. Phys.* **129** 074704
[24]    Peng X H and Velasquez S 2011 *Appl. Phys. Lett.* **98** 023112
[25]    Neek-Amal M and Peeters F M 2011 *Journal of Physics-Condensed Matter* **23** 243201
[26]    Yu D and Liu F 2007 *Nano Lett.* **7** 3046
[27]    Jung N, Crowther A C, Kim N, Kim P and Brus L 2010 *ACS Nano* **4** 7005
[28]    Rojas M I and Leiva E P M 2007 *Phys. Rev. B* **76** 155415
[29]    Bermudez V M and Robinson J T 2011 *Langmuir* **27** 11026
[30]    Leenaerts O, Peelaers H, Hernández-Nieves A D, Partoens B and Peeters F M 2010 *Phys. Rev. B* **82** 195436
[31]    Sofo J O, Chaudhari A S and Barber G D 2007 *Phys. Rev. B* **75** 153401

[32] Boukhvalov D W and Katsnelson M I 2009 *Journal of Physics-Condensed Matter* **21** 344205
[33] Kohn W and Sham L J 1965 *Physical Review* **140** A1133
[34] Kresse G and Furthmuller J 1996 *Comput. Mater. Sci.* **6** 15
[35] Kresse G and Furthmuller J 1996 *Phys. Rev. B* **54** 11169
[36] Perdew J P, Chevary J A, Vosko S H, Jackson K A, Pederson M R, Singh D J and Fiolhais C 1992 *Phys. Rev. B* **46** 6671
[37] Oshima C and Nagashima A 1997 *Journal of Physics: Condensed Matter* **9** 1
[38] Vanderbilt D 1990 *Phys. Rev. B* **41** 7892
[39] Monkhorst H J and Pack J D 1976 *Phys. Rev. B* **13** 5188
[40] Nakada K, Fujita M, Dresselhaus G and Dresselhaus M S 1996 *Phys. Rev. B* **54** 17954
[41] Son Y W, Cohen M L and Louie S G 2006 *Phys. Rev. Lett.* **97** 216803
[42] Shan B and Cho K 2005 *Phys. Rev. Lett.* **94** 236602
[43] Gong H R, Nishi Y and Cho K 2007 *Appl. Phys. Lett.* **91** 242105
[44] Bagus P S, Illas F, Pacchioni G and Parmigiani F 1999 *Journal of Electron Spectroscopy and Related Phenomena* **100** 215
[45] Richter B, Kuhlenbeck H, Freund H J and Bagus P S 2004 *Phys. Rev. Lett.* **93** 026805
[46] Grant R W, Waldrop J R, Kraut E A and Harrison W A 1990 *Journal of Vacuum Science & Technology B* **8** 736
[47] Hofmann O T, Egger D A and Zojer E 2010 *Nano Lett.* **10** 4369
[48] Rusu P C, Giovannetti G, Weijtens C, Coehoorn R and Brocks G 2009 *Journal of Physical Chemistry C* **113** 9974
[49] Crispin X, Geskin V, Crispin A, Cornil J, Lazzaroni R, Salaneck W R and Brédas J-L 2002 *Journal of the American Chemical Society* **124** 8131
[50] Fulton C C, Lucovsky G and Nemanich R J 2004 *Appl. Phys. Lett.* **84** 580

**Table caption**

Table 1 The selected bond lengths (in unit of angstrom) in the relaxed and -12% strained AGNRs with a width of 14 C atoms. The number notation of atoms is indicated in Fig. 5.

**Figure captions**

FIG. 1. The snapshots of AGNRs with edge carbon atoms passivated by (a) H and (b) O. The dashed rectangles indicate unit cells. The width of a ribbon is closely related to the number of carbon atoms in the *y* direction. The pictures show the AGNRs with a width of 14 C atoms.

FIG. 2. The work function of the AGNRs with a width of 14 C atoms as a function of uniaxial strain. The vertical lines indicate the strains at which the structural/electronic transitions occur.

FIG. 3. (a) The variation of Fermi and vacuum levels with strain; (b) the normalized core level shift relative to the Fermi energy and vacuum level under strain for the edge-H passivated AGNR with a width of 14 C atoms.

FIG. 4. The band structure of the edge-O passivated AGNR with a width of 14 C atoms under different values of uniaxial strain. The energies are referenced to the vacuum level. The band gap experiences a transition from direct to indirect at +4% strain, and shrinks to zero at +8% strain and beyond. A structural transformation occurs at -12% strain, producing a largely deviated band structure.

FIG. 5. The structures of the (a) relaxed and (b) -12% strained AGNRs with a width of 14 C atoms in two adjacent simulation cells. Note that a structural transformation occurs at -12% strain. The horizontal C-C pairs, such as C3-C4, C5-C6, C9-C10, C11-C12, and C13-C14, have totally different bond distances compared to that of the relaxed ribbon. And edge defects such as carbon quadrilaterals form in the -12% strained ribbon.

FIG. 6. The AGNRs surface decorated by H atoms. (a) no H atoms on the surface; (b) 4 H atoms on one side; (c) 8 H atoms on one side; (d) 8 H atoms on both sides (4 H each side). Each addition of one H corresponds to an increment of 3.57% surface coverage. The top and bottom rows represent different views indicated by the coordinate axes.

FIG. 7. (a) The relaxed lattice constant and (b) the bending curvature of the AGNRs with different surface functional species.

FIG. 8. The band structure and the corresponding density of states of the AGNRs with different surface functional species decorated on one side of the ribbon surface. The starting ribbon is the edge-H passivated AGNR with a width of 14 C atoms in which no surface species were decorated. The notation "surface 2H" means two H atoms were decorated on the ribbon surface.

FIG. 9. The work function of the AGNRs with different surface functional species.

| Bond | Length for relaxed AGNR (Å) | Length for -12% strained AGNR (Å) | Difference (Å) |
|---|---|---|---|
| C3-C4 | 1.42 | 2.30 | 0.88 |
| C3'-C4 | 2.68 | 1.31 | -1.37 |
| C5-C6 | 2.73 | 1.29 | -1.44 |
| C5'-C6 | 1.37 | 2.31 | 0.95 |
| C1-C2 | 2.47 | 2.20 | -0.27 |
| C1'-C2 | 1.63 | 1.40 | -0.23 |
| C1-C3 | 1.39 | 1.47 | 0.09 |
| C3-C5 | 1.40 | 1.35 | -0.05 |
| C1-O7 | 1.50 | 1.35 | -0.15 |
| C2-O7 | 1.50 | 1.35 | -0.15 |

**Table 1**

(a) Edge-H passivated AGNR 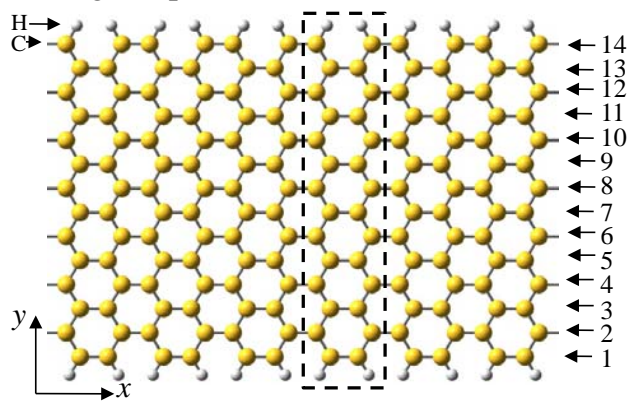

(b) Edge-O passivated AGNR 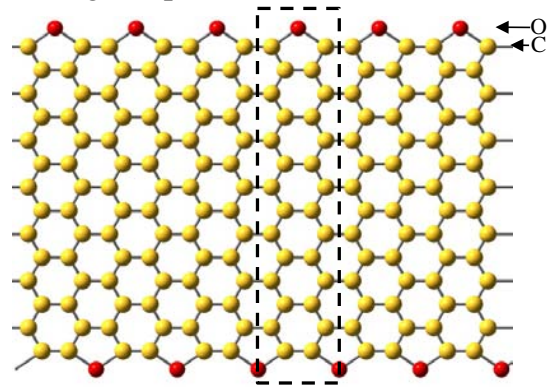

**FIG. 1**

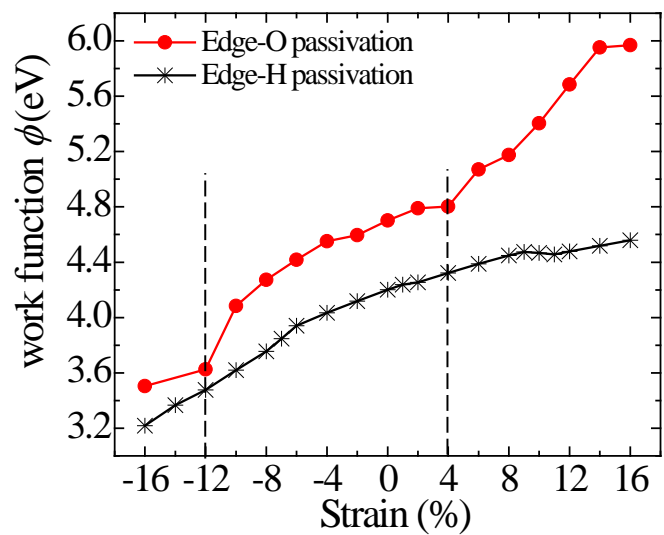

**FIG. 2**

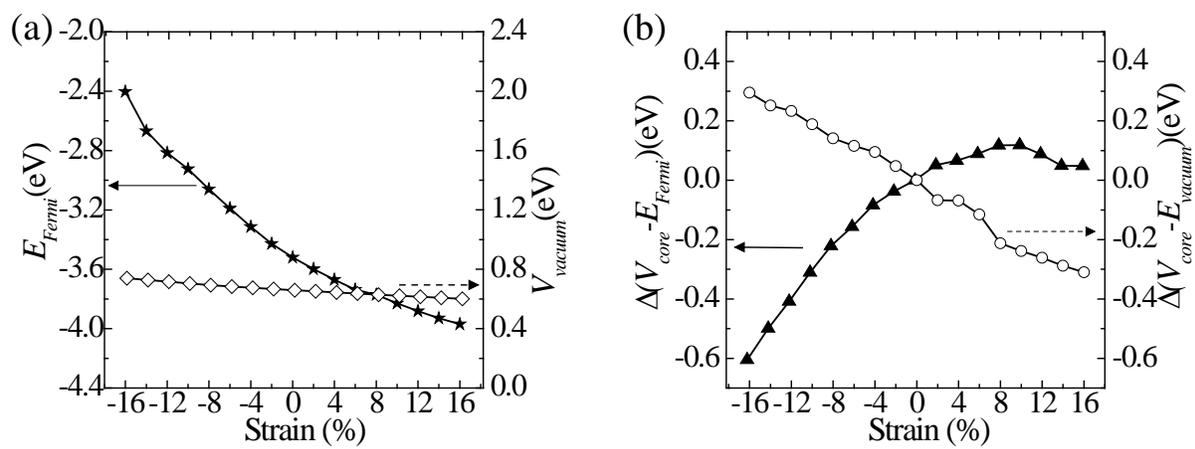

**FIG. 3**

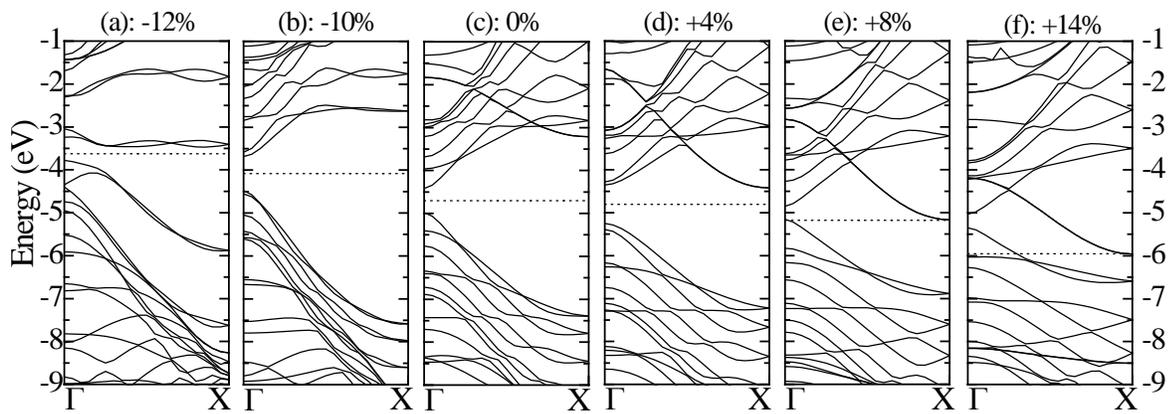

**FIG. 4**

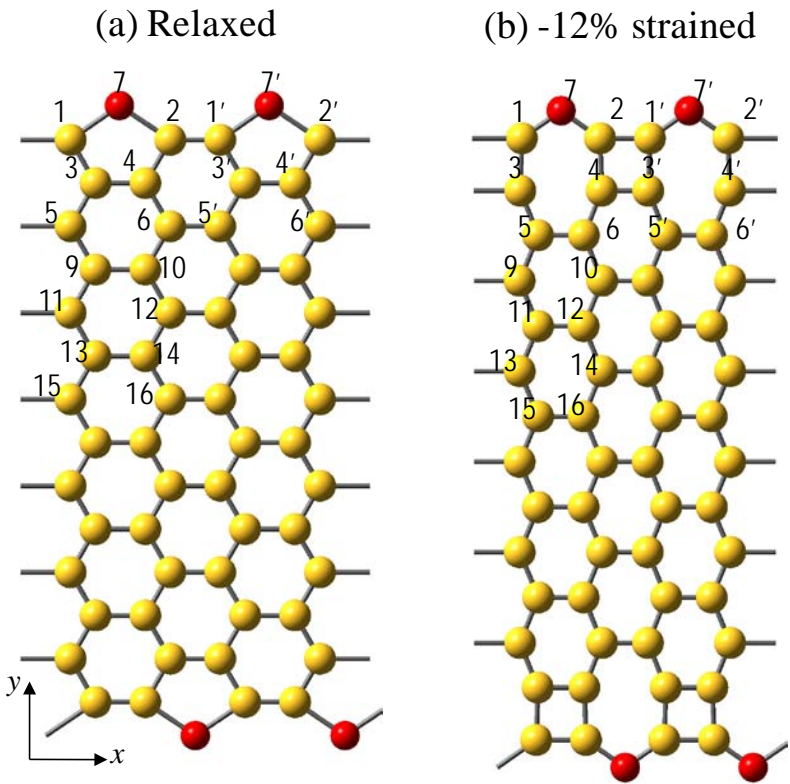

**FIG. 5**

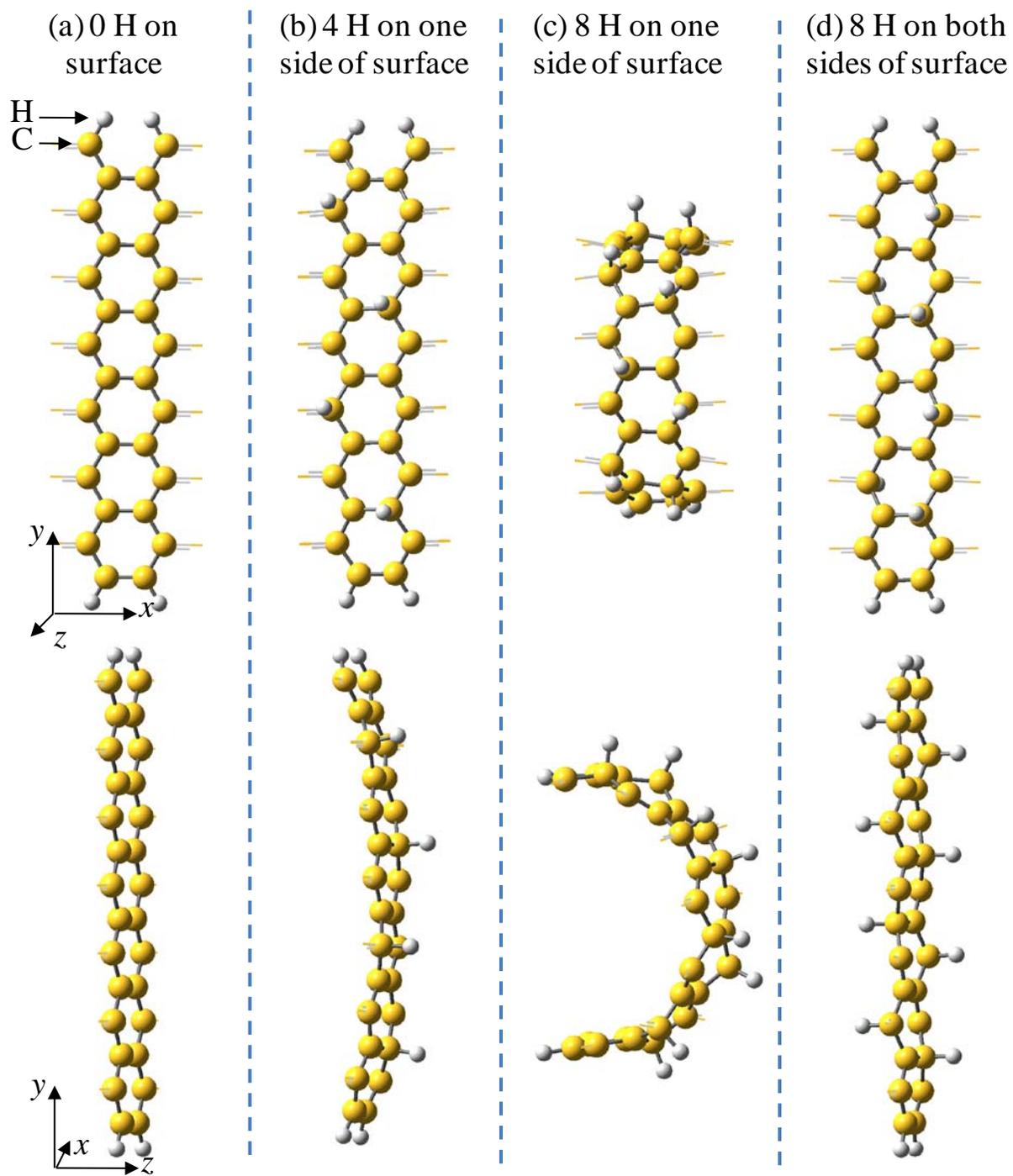

**FIG. 6**

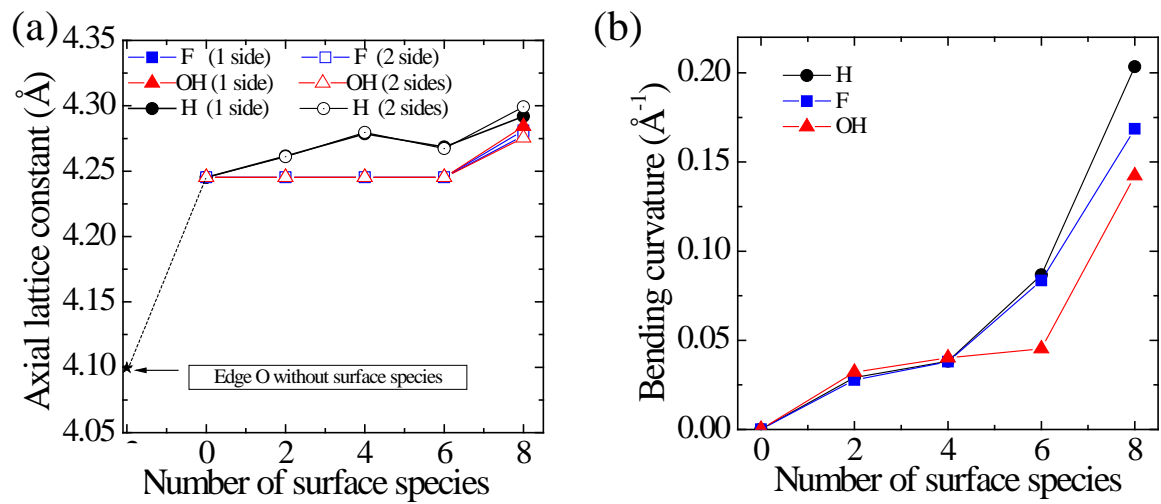

**FIG. 7**

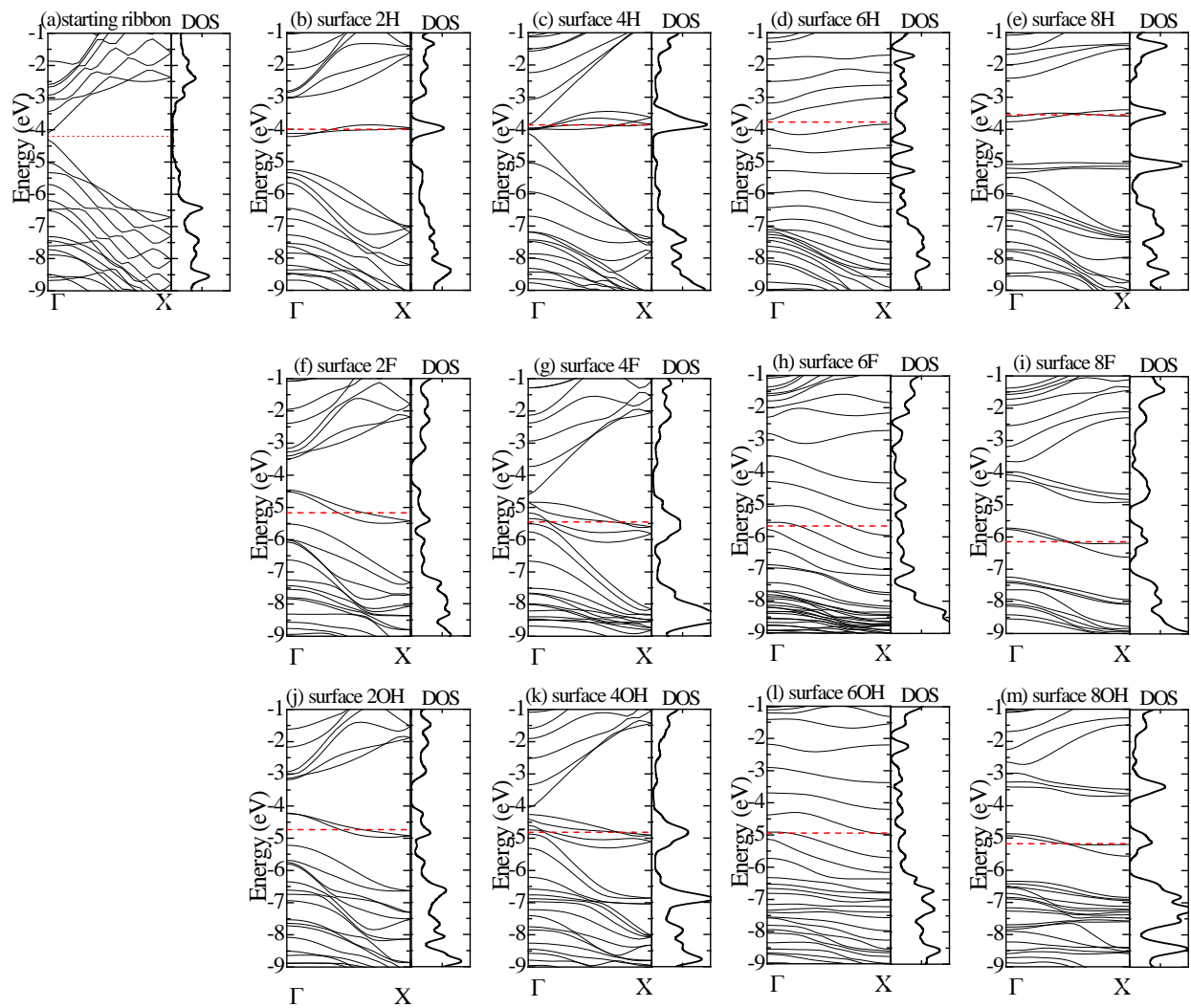

**FIG. 8**

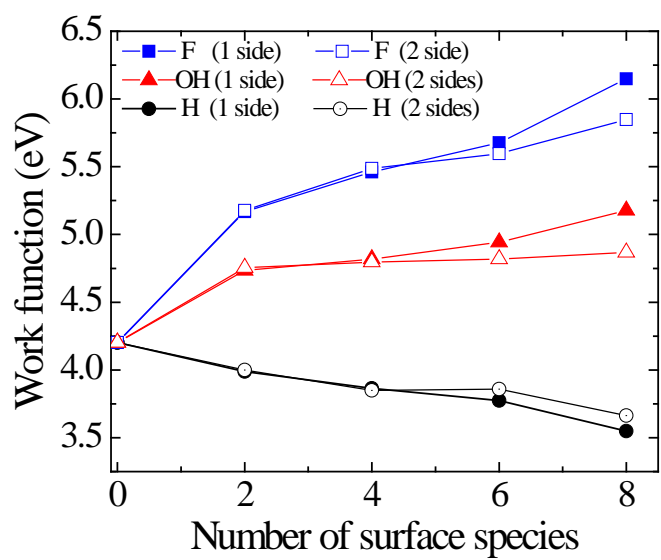

**FIG. 9**